\documentclass[conference]{IEEEtran}
\usepackage{cite}
\usepackage{amsmath,amssymb,amsfonts}
\usepackage{balance}
\usepackage{algorithmic}
\usepackage{graphicx}
\usepackage{pgfplots}
\usepackage{textcomp}
\usepackage{xcolor}
\usepackage{caption}
\usepackage{subcaption}
\def\BibTeX{{\rm B\kern-.05em{\sc i\kern-.025em b}\kern-.08em
    T\kern-.1667em\lower.7ex\hbox{E}\kern-.125emX}}
\usepackage{url}
\usepackage{glossaries}
\newacronym{fpga}{FPGA}{{Field-Programmable Gate Array}}
\newacronym{cots}{COTS}{{Commercial Off-The-Shelf}}
\newacronym{os}{OS}{{Operating System}}
\newacronym{rt}{RT}{{Real-Time}}
\newacronym{rtos}{RTOS}{{Real-Time Operating System}}
\newacronym{irq}{IRQ}{{Interrupt Request}}
\newacronym{isr}{ISR}{{Interrupt Service Routine}}
\newacronym{ist}{IST}{{Interrupt Service Task}}
\newacronym{iot}{IoT}{{Internet of Things}}
\newacronym{nic}{NIC}{{Network Interface Controller}}
\newacronym{mcu}{MCU}{{Microcontroller Unit}}

\makeglossaries

\pgfplotsset{compat=1.8}
\begin{document}

\title{Interrupting Real-Time IoT Tasks:\\ How Bad Can It Be to Connect Your Critical Embedded System to the Internet?}

\author{

\IEEEauthorblockN{Ilja Behnke\IEEEauthorrefmark{1}, Lukas Pirl\IEEEauthorrefmark{3}, Lauritz Thamsen\IEEEauthorrefmark{1}, Robert Danicki\IEEEauthorrefmark{1}, Andreas Polze\IEEEauthorrefmark{3}, and Odej Kao\IEEEauthorrefmark{1}}

\IEEEauthorblockA{
\IEEEauthorrefmark{1}
Technische Universität Berlin, Germany, i.behnke@tu-berlin.de\\
}
\IEEEauthorblockA{
\IEEEauthorrefmark{3}
Hasso Plattner Institute, Germany, \{first.last\}@hpi.de\\
}
}

\IEEEoverridecommandlockouts
\IEEEpubid{\makebox[\columnwidth]{978-1-7281-9829-3/20/\$31.00~\copyright2018 IEEE \hfill} \hspace{\columnsep}\makebox[\columnwidth]{ }}

\maketitle

\IEEEpubidadjcol

\begin{abstract}
Embedded systems have been used to control physical environments for decades. Usually, such use cases require low latencies between commands and actions as well as a high predictability of the expected worst-case delay. To achieve this on small, low-powered microcontrollers, Real-Time Operating Systems (RTOSs) are used to manage the different tasks on these machines as deterministically as possible. However, with the advent of the Internet of Things (IoT) in industrial applications, the same embedded systems are now equipped with networking capabilities, possibly endangering critical real-time systems through an open gate to interrupts.

This paper presents our initial study of the impact network connections can have on real-time embedded systems. Specifically, we look at three aspects: The impact of network-generated interrupts, the overhead of the related networking tasks, and the feasibility of sharing computing resources between networking and real-time tasks. We conducted experiments on two setups: One treating NICs and drivers as black boxes and one simulating network interrupts on the machines. The preliminary results show that a critical task performance loss of up to 6.67\% per received packet per second could be induced where lateness impacts of 1\% per packet per second can be attributed exclusively to ISR-generated delays.

\end{abstract}

\begin{IEEEkeywords}
Internet of Things, Real-time, Interrupts, Cyber Physical Systems, Embedded Systems
\end{IEEEkeywords}

\section{Introduction}
Embedded devices that control machines in the physical world have been part of industrial processes as well as home and automotive appliances for decades \cite{microcontrollers, robots, indusTempMonit}. In contrast to general-purpose
computing, these devices need to fulfill timing constraints.
To this end, \glspl{rtos} are used, which are lightweight and make guarantees towards the timing predictability
of tasks \cite{rtos}.
Usually, a preemptive task scheduler allows to configure different priorities for different concurrently active tasks, so that the most time-critical tasks always take precedence over less critical ones. 

\glspl{irq} are generated by the hardware and inevitable for systems to function.
At the same time, they introduce a level of unpredictability to the process flow.
Since the corresponding \glspl{isr} are handled by the processor, the scheduler of an \gls{os} has no control over their execution.
However, by keeping the execution times of \glspl{isr} minimal and considering worst-case scenarios during the development, most traditional embedded systems can handle \glspl{irq} without missing deadlines.
Yet, in the past, the environment controlled by embedded systems tended to be self-contained, the number of environmentally-triggered \glspl{irq} was typically small, and their impact, therefore, predictable.
With the advent of the \gls{iot} in industrial applications this premise has changed.
\gls{iot} microcontrollers come with built-in network chips and are increasingly often network-connected for the sake of remote control, monitoring, and maintenance \cite{iiot_mcu}.
\gls{iot} networks are, however, open by design and thus less policed \cite{zhang15communication}.
Especially for critical real-time tasks on networked embedded microcontrollers, this is a threat:
The embedded systems have to handle the additional resource consumption of the non-critical networking tasks and the necessary \gls{nic} introduces a new source of unpredictability as incoming packets trigger \glspl{irq} that disturb the flow of scheduled tasks \cite{hip}. This might lead to a critical load of interrupts and triggered network tasks in the \gls{rtos}, invalidating real-time guarantees and thereby lowering the system's dependability.

This paper analyzes the impact of network loads on critical real-time tasks running on state-of-the-art microcontrollers with modern \glspl{rtos} used in the \gls{iot}. In an initial study, we evaluate timing measurements of critical tasks on
microcontrollers running vendor-supplied
\glspl{rtos}, network drivers, and network stack tasks under different network-triggered \gls{irq} loads. To expose any existing mitigation in the hardware and closed source drivers, a pseudo network driver is designed to serve as a second \gls{irq} source.
Measurements are taken and compared between real and pseudo network packet processing. Building on our methodology, we perform the following contributions:
\begin{itemize}
\item Measurement of ISR-induced delay to real-time tasks.
\item Evaluation of overhead induced by networking tasks under different network loads.
\item Preliminary analysis of the feasibility of IoT programming frameworks and IP networking in real-time scenarios.
\end{itemize}

\emph{Outline}. The remainder of the paper is structured as follows.
Section~\ref{sec:problem} presents the problem statement.
Section~\ref{sec:methodology} introduces our evaluation methodology.
Section~\ref{sec:evaluation} presents our empirical results with two different setups.
Section~\ref{sec:discussion} discusses the results.
Section~\ref{sec:related} describes the related work, while Section~\ref{sec:conclusion} concludes the paper.

\section{Problem Statement}
\label{sec:problem}
As motivated, the inclusion of network controllers to embedded real-time systems introduces an unpredictable source of interrupts. The goal of this work is the analysis of the impact of these interrupts in IoT environments. This includes the execution of ISRs, network drivers, and network stack tasks, as well as the robustness of the entire system under high network loads. This section provides the scope and assumptions of our work.

\subsection{IoT Environments}
Available IoT devices are microcontrollers possessing the means to wirelessly connect to networks and handle a multitude of different network protocols comparable to the network stack implementations in general computing. At the same time, these systems are designed to be deployed in untrusted environments, more or less directly connected to the internet \cite{iotnetstack}. Packet floods generated by faults and security breaches furthermore lay open potentially critical systems that are deployed in secluded networks. 
Due to these circumstances, we consider networked embedded systems of increased vulnerability while controlling critical systems with real-time constraints \cite{bellekens2015cyber}.

The high popularity of the IoT led to the mass production of IP capable devices that are cheaply available. To make the somewhat complex programming for embedded systems more available in the IoT context, devices come with programming frameworks masking low-level software like network stacks and drivers. 
While the investigated modules can technically also be programmed without their respective frameworks, the immense workload of the implementation and inclusion of the network driver and stack tasks makes this unfeasible in most cases. An incorporation of the built-in Wi-Fi chips into fully manageable and transparent real-time systems is hindered by the unavailability of driver source code. 

\subsection{Design and Development for IoT MCUs}
During the design and development of real-time systems, the developer is responsible for making sure that any deadlines are met in the worst-case scenario. In case of externally caused interrupts, it is therefore necessary for the developer to incorporate the maximum frequency of interrupts and their impact into the design of the system. Network-generated interrupts however leave the developer only with few choices like turning off interrupts or reserving a physical core for networking tasks \cite{hermes}. While both actions might solve the problem of interrupt floods over the network, the loss of a core (if available) and access to the embedded system from outside might not be feasible.

Additionally, the limited access to low-level functions leads to a loss of control over the real-time system, which should generally be designed holistically. 

\subsection{Assumptions}
Concluding, we make the following assumptions for our analysis. 
\begin{itemize}
\item \textit{A1 - Unreliable networks}: IoT devices are commonly deployed in untrustworthy networks and/or might be subject to network faults.
\item \textit{A2 - WiFi}: While the issue of unpredictable interrupts is the same across network interfaces, driver availability and specific processing costs differ. Wi-Fi connections belong to the most commonly used interfaces, since no modifications to available hardware has to be made and the technology is mature.
\item \textit{A3 - Shared Resources}: Networking tasks, drivers, and application tasks might run on the same MCU core.
\item \textit{A4 - Device-specific frameworks}: To implement real-time systems with Wi-Fi capabilities on IoT devices, device-specific frameworks are used.
\end{itemize}

The design of our experiment setup is deviated from these assumptions. 

\section{Methodology}
\label{sec:methodology}

A certain impact of networking tasks and interrupts on \gls{mcu} utilization and timing predictability in real-time embedded systems seems inherent. With the experimental methodology and setup we aim to analyze this impact both qualitatively and quantitatively on two current \gls{cots} IoT devices.

\subsection{Experiment Design}

For the quantitative analysis we designed two sets of experiments observing timing metrics of a periodic task. The experiments are run under different network loads with two regarded interrupt setups (cf. Section \ref{sec:intrsetups}). The experiment permutations were additionally adapted to take into account the differences between priorities of preexisting network tasks.

\subsubsection*{Observed Metrics}
We defined two sets of experiments observing different metrics under changing network loads.

\subsubsection{Lateness}
\begin{figure}[ht]
\centering
\includegraphics[scale=1]{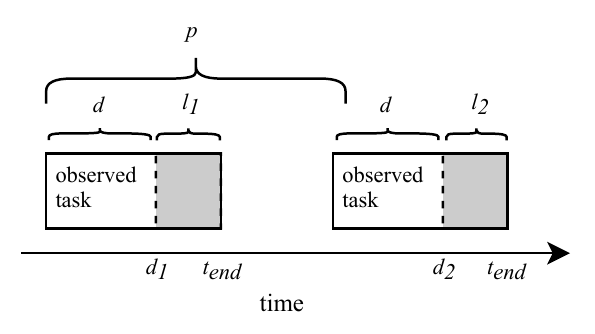}
\caption{Lateness measurements.}
\label{fig:lateness}
\end{figure}

The critical task we observe is periodically called by a timer with period \textit{p} and a deadline \textit{d}. We define the lateness \textit{l} as the time the task takes longer to finish than its deadline allows, hence $ l = t_{end} - d $ where $t_{end}$ is the time at which one process cycle of the task has finished as depicted in Figure \ref{fig:lateness}. 
Once we reach a load at which the task misses its deadline, the lateness will start to accumulate over iterations. We present the accumulated lateness per second.

\subsubsection{Relative MPU utilization}
In the second set of experiments, the observed task runs in a closed loop for a duration of $\Delta t$ in an interval \textit{i}. By measuring the number of cycles the critical task finishes in one $\Delta t$ under different network loads we can make out the resource utilization of networking operations depending on the number of received packets.

\subsection{Test Environment}
The experiments were performed on two widely used IoT development boards in similar performance and price ranges. One equipped with a dual-core ESP32 chip at 160~MHz, the other with a Particle Photon (P0) single-core ARM Cortex M3 at 120~MHz. While both devices have ports of the FreeRTOS operating system
, the vendor-provided programming frameworks differ significantly in terms of programmability and port specific implementations. Both devices are programmed using their respective frameworks
. While the frameworks themselves and operating systems are open source, some low-level software such as Wi-Fi drivers are only available as binary objects. 
While the  differences between frameworks and used processing chips limit the degree of direct comparability, they allow us to evaluate the possibilities of development inside the frameworks' constraints. This way, all results are realistic for the systems tested under their established workflows.

\subsection{Task Priorities}
\label{sec:priorities}
To keep the connections alive, driver and networking tasks have to be kept running on the devices. These are called by NIC-triggered ISRs but can be preempted by a higher prioritized task on the same core.
To evaluate the different shares of delays caused by ISRs and networking tasks, experiments are repeated with observed task priorities chosen above, equal to, and below the driver. The default networking task and driver priorities are specified by the frameworks as depicted in Table \ref{tab:priorities}. Wi-Fi driver priorities cannot be changed.

\begin{table}[h!]
\caption{Networking task priorities module in firmwares}
\label{tab:priorities}
\centering
\begin{tabular}{r|c|c|l}
\hline
\textbf{prio} & \textbf{ESP32} & \textbf{P0} & \textbf{prio}  \\
\hline
\textit{24} & - & Wi-Fi driver & \textit{9} \\
\textit{23} &Wi-Fi driver & network (high) & \textit{8} \\
\textit{22-19} & - & network (low) & \textit{7} \\
\textit{18} & network & - & \textit{6-1} \\
\hline
\end{tabular}
\end{table}

\subsection{Interrupt Workload}
\label{sec:intrsetups}
To analyze the impact on lateness and computing resource utilization, the devices under test are put under different network loads.

\subsubsection*{Generation}
One of the main difficulties when investigating the process flow a received packet triggers is the unavailability of large parts of the driver source code. Quantitatively analyzing the number of times an ISR is effectively called is therefore difficult to realize. To be able to compare the relative impact of ISRs and low-level packet processing, two interrupt setups are run on the devices.

\subsubsection{Real Network Packets}
In the first setup, network packets are sent to the devices over a Wi-Fi connection 
and are handled by the framework-supplied driver and networking tasks (Figure \ref{fig:setup_net}). We use the second core of the ESP32 to run a minimally configured UDP server to measure its impact. 
As the P0 does not have a dual-core processor, no UDP server was run and interrupts were generated externally.

\subsubsection{Simulated Packets}
\begin{figure}[hb]
\centering
\begin{subfigure}{.45\columnwidth}
	\centering
	\includegraphics[scale=1]{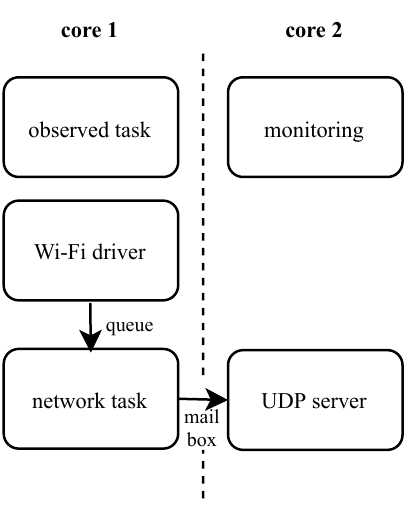}
	\caption{Wi-Fi driver}
	\label{fig:setup_net}
\end{subfigure}
\hfill
\begin{subfigure}{.45\columnwidth}
	\includegraphics[scale=1]{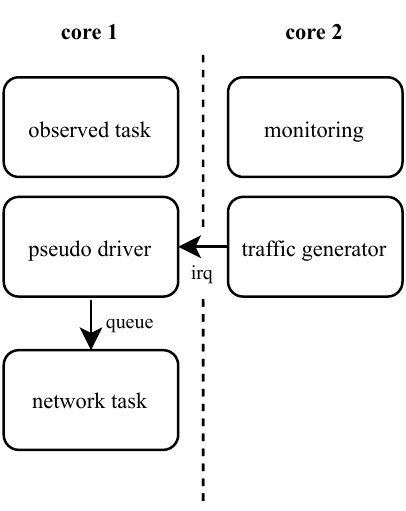}
	\caption{network simulator}
	\label{fig:setup_sim}
\end{subfigure}
\caption{Experiment setups on ESP32.}
\label{fig:setups}
\end{figure}

Secondly, we implemented an analogous task flow to perform Wi-Fi driver independent experiments (Figure \ref{fig:setup_sim}). A networking task simulation largely performs the same actions a packet received over a network triggers: Upon registering the interrupt which in this case is triggered via an input pin, a short ISR is called that preempts the currently running process to copy a packet descriptor to a FreeRTOS queue. A pseudo driver task waits for packets in this queue and unblocks when it is not empty to process the entries.

\subsubsection*{Traffic Loads}
Changes in network load directly affect the number of ISR, driver, and network stack calls in the operating system. Depending on the system and interrupt type, network loads were increased in steps of 10, 100, or 1,000 packets per second and held to take measurements. Additionally, experiments were conducted with traffic bursts of 120,000 packets per second for one second.

\pgfplotsset{every axis/.append style={very thick}, legend style={font=\small}, label style={font=\small}, tick label style={font=\small}}

\section{Experimental Results}
\label{sec:evaluation}
This section presents our preliminary empirical results of how interrupt loads impact the lateness of critical tasks and utilization of CPUs.

\begin{figure*}
\centering
\begin{subfigure}{.3\textwidth}
	\begin{tikzpicture}
		\begin{axis}[width=\textwidth,
			xmin=0,
			xmax=125000,
			x tick label style={/pgf/number format/1000 sep=},
			xlabel={packets / s},
			y tick label style={/pgf/number format/1000 sep=},
			ylabel={lateness [ms]},
			legend pos=north east,
			grid=major,
			grid style={dashed},
			no markers
			]

			\addplot+ [dotted, color=teal] table[x=sent_low, y=late_low, col sep=comma]{data/plot1.csv};
		 	\addlegendentry{p-}
		 	\addplot+ [dashed, color=purple] table[x=sent, y=late_high, col sep=comma]{data/plot1.csv};
		 	\addlegendentry{p+}
		\end{axis}
	\end{tikzpicture}
	\caption{ESP32, simulated network traffic}
	\label{fig:plot1}
\end{subfigure}
\hfill
\begin{subfigure}{.3\textwidth}
	\begin{tikzpicture}
		\begin{axis}[width=\textwidth,
			xmin=0,
			xmax=4200,
			ymax=1200,
			x tick label style={/pgf/number format/1000 sep=},
			xlabel={packets / s},
			y tick label style={/pgf/number format/1000 sep=},
			ylabel style = {align=center},
			ylabel={lateness [ms]},
			legend style={cells={align=left}},
			legend style={font=\tiny},
			legend pos=north west,
			grid=major,
			grid style={dashed},
			no markers
			]

			\addplot+ [dotted, color=teal] table[x=sent, y=late_low, col sep=comma]{data/plot5.csv};
		 	\addlegendentry{p- bound}
		 	\addplot+ [dashed, color=purple] table[x=sent, y=late_high, col sep=comma]{data/plot5.csv};
		 	\addlegendentry{p+}
		 	\addplot+ [loosely dashdotted, color=gray] table[x=sent, y=late_low_noudp, col sep=comma]{data/plot5.csv};
		 	\addlegendentry{p-}
		\end{axis}

	\end{tikzpicture}
	\caption{ESP32, real IP packets}
	\label{fig:plot5}
\end{subfigure}
\hfill
\begin{subfigure}{.3\textwidth}
	\begin{tikzpicture}
		\begin{axis}[width=\textwidth,
			xmin=0,
			xmax=1020,
			x tick label style={/pgf/number format/1000 sep=},
			xlabel={packets / s},
			y tick label style={/pgf/number format/1000 sep=},
			ylabel={lateness [ms]},
			legend pos=north west,
			grid=major,
			grid style={dashed},
			no markers
			]

			\addplot+ [dotted, color=teal] table[x=sent, y=late_low, col sep=comma]{data/plot6.csv};
		 	\addlegendentry{p-}
		 	\addplot+ [dashed, color=purple] table[x=sent, y=late_high, col sep=comma]{data/plot6.csv};
		 	\addlegendentry{p+}
		\end{axis}
	\end{tikzpicture}
	\caption{P0, real IP packets}
	\label{fig:plot6}
\end{subfigure}
\caption{Lateness experiment results with observed task priorities under networking tasks (p-) and critical (p+).}
\label{fig:lateplots}
\end{figure*}
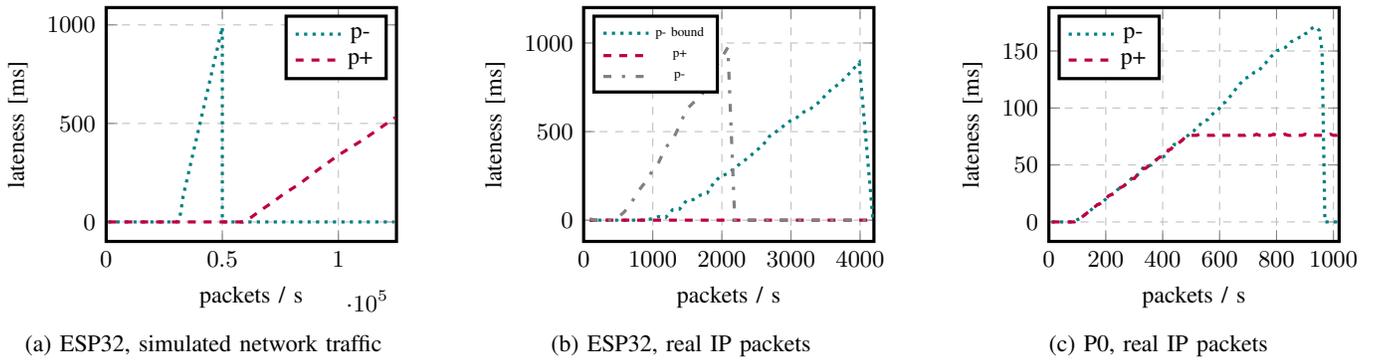

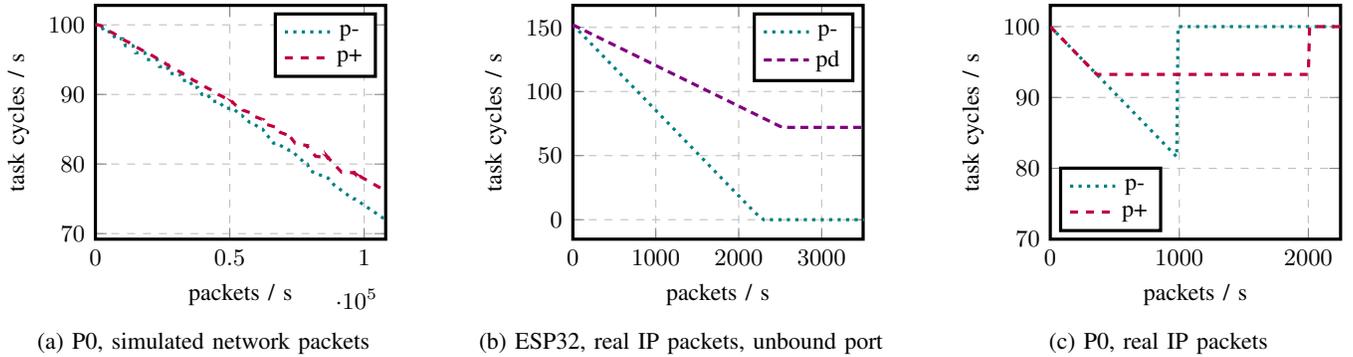
\begin{figure*}
\centering
\begin{subfigure}{.3\textwidth}
	\begin{tikzpicture}
		\begin{axis}[width=\textwidth,
			xmin=0,
			xmax=108000,
			x tick label style={/pgf/number format/1000 sep=},
			xlabel={packets / s},
			y tick label style={/pgf/number format/1000 sep=},
			ylabel={task cycles / s},
			legend pos=north east,
			grid=major,
			grid style={dashed},
			no markers
			]

			\addplot+ [dotted, color=teal] table[x=rcv_low, y=cc_low, col sep=comma]{data/plot8.csv};
		 	\addlegendentry{p-}
		 	\addplot+ [dashed, color=purple] table[x=rcv_high, y=cc_high, col sep=comma]{data/plot8.csv};
		 	\addlegendentry{p+}
		\end{axis}
	\end{tikzpicture}
	\caption{P0, simulated network packets}
	\label{fig:plot8}
\end{subfigure}
\hfill
\begin{subfigure}{.3\textwidth}
	\begin{tikzpicture}
		\begin{axis}[width=\textwidth,
			xmin=0,
			xmax=3500,
			x tick label style={/pgf/number format/1000 sep=},
			xlabel={packets / s},
			y tick label style={/pgf/number format/1000 sep=},
			ylabel={task cycles / s},
			legend pos=north east,
			grid=major,
			grid style={dashed},
			no markers
			]

			\addplot+ [dotted, color=teal] table[x=sent, y=cc_low, col sep=comma]{data/plot10.csv};
		 	\addlegendentry{p-}
		 	\addplot+ [densely dashed, color=violet] table[x=sent, y=cc_high, col sep=comma] {data/plot10.csv};
		 	\addlegendentry{pd}
		\end{axis}
	\end{tikzpicture}
	\caption{ESP32, real IP packets, unbound port}
	\label{fig:plot10}
\end{subfigure}
\hfill
\begin{subfigure}{.3\textwidth}
	\begin{tikzpicture}
		\begin{axis}[width=\textwidth,
			ymin=70,
			xmin=0,
			xmax=2250,
			x tick label style={/pgf/number format/1000 sep=},
			xlabel={packets / s},
			y tick label style={/pgf/number format/1000 sep=},
			ylabel={task cycles / s},
			legend pos=south west,
			grid=major,
			grid style={dashed},
			no markers
			]

			\addplot+ [dotted, color=teal] table[x=sent, y=cc_low, col sep=comma]{data/plot11.csv};
		 	\addlegendentry{p-}
		 	\addplot+ [dashed, color=purple] table[x=sent, y=cc_high, col sep=comma] {data/plot11.csv};
		 	\addlegendentry{p+}
		\end{axis}
	\end{tikzpicture}
	\caption{P0, real IP packets}
	\label{fig:plot11}
\end{subfigure}
\caption{Task cycle counter experiment results with observed task priorities under networking tasks (p-), critical (p+), and equal with wi-fi driver (pd).}
\label{fig:ccplots}
\end{figure*}

\subsection{Lateness Experiments}
\subsubsection*{ESP32}
The first group of experiments was conducted to measure the lateness of the observed task under rising packet loads. Figures \ref{fig:plot1} and \ref{fig:plot5} contain the lateness results under simulated and real network traffics on the ESP32. Both show a linear increase in lateness with rising packet load once lateness occurs for priorities chosen below the driver priory with real IP packet impact reaching 50\% lateness increase per packet per second. 
The differences between priorities correspond to the preemption of the fixed networking tasks once the MCUs are fully utilized. When the observed task has a lower priority than the networking tasks, it starves once too many packets arrive. When the observed task has a critical priority, it in turn starves the networking tasks. As can be seen this results in no impact, suggesting that no ISRs are triggered by the NIC in this case. 

Figure \ref{fig:plot5} also shows that the impact of incoming packets is a lot higher when addressed to a port that is not listened on. This might be due to the partial deactivation of networking tasks when UDP buffers in the network stack are full.

Lateness results under traffic bursts do not differ from continuous loads and no other influence to the systems could be seen. Further burst results are therefor omitted from this work as this was true for all burst experiments.

\subsubsection*{P0}

Figure \ref{fig:plot6} contains the analogous results of the experiments on the P0. The results of the simulated approach are very similar to the ESP32 equivalent, with the difference in slope explainable by platform specific pseudo driver implementations. The results from the experiments using real network packets show an impact slope of 2.2\%. The results also show issue of the network driver residing on the highest priority level. Once the packet load is high enough that the network driver needs half of the computing resources for itself, the operating system's scheduler distributes the processing resources equally between the driver and the observed (critical) task. 
In contrast to ESP32 task, the observed task here does not starve when of lower priority. Before this can happen, the Wi-Fi task crashes at 980 packets per second. 

\subsection{Utilization Experiments}

Figure \ref{fig:ccplots} contains the parallel results for real network packets and software. Task priorities of the presented results were chosen equal to and lower than the Wi-Fi driver. Task performance decreases by 2.15\% and 3.8\% respectively. When receiving packets on an unused port the impact is again higher with performance decreases of 3.17\% and 6.67\% respectively.

Figure \ref{fig:plot8} contains the results for the network driver simulation on the P0, which are comparable to the lateness results. With real network packets the system crashes at 980 packets per second when giving the observed task a lower priority than the Wi-Fi driver and at 2,000 packets per second with the same priority. Task performance decreases by 2\% per packet per second until equal resource utilization with the driver (same priority) or throughout gradual extrusion by it (lower priority).

\section{Discussion}
This section discusses the findings from the experiments and the feasibility of IP networking on \glspl{mcu}.

\label{sec:discussion}
\subsection{Insights from the Experiment Results}

The evaluation results show that the performance and real-timeness, is directly dependent on incoming IP network load. A high overhead generated by the receiving of packets can be seen in continuous floods as well as short transmission bursts. The Wi-Fi driver and network stack utilize any compute resources they need to handle packets unless preempted. Yet, the results also show that NIC-triggered ISRs have no impact on the observed systems. This observation is made when incoming packets are not handled due to the driver being preempted by a task of critical priority. 

To mitigate the observed breaking of real-time guarantees on the tested devices, developers still have some options. The ESP IDF provides one priority level above the Wi-Fi task's priority. Running a critical task here will preempt the driver. For critical code sections it is also possible to disable all or only the Wi-Fi interrupt. This is the only option under Particle's DeviceOS.

\subsection{Feasibility of IP Networking}
The main take away is that it is necessary to fit all mission-critical operations into a critical task that is independent of any signals received over the network. Hence, performing command and control operations over the network cannot be an option for real-time systems. 
This however invalidates most IoT use-cases.

Network driver tasks, which are responsible for a large share of the impact on timing predictability and hence lateness, are very highly prioritized in the tested systems' frameworks. 
This highly limits the margin critical tasks have. While short and independent tasks and code blocks can be executed in a safe manner, this is also where it ends.

Network driver tasks are currently given a priority level in RTOSs that is not suitable for critical real-time systems. Using network connectivity only for monitoring timing independent tasks might still be an option when one is ready to provide the resources and prioritize the necessary tasks appropriately.

\subsection{Networking Software Robustness}
Current IoT device programming frameworks seem to be designed for certain connectivity guarantees rather than real-timeness of critical applications. However, this also does not hold for the evaluated systems as can be seen from the results. Under the conditions of the test setup, even low traffic loads lead to Wi-Fi driver crashes and, depending on the framework configuration, to system halts. This could be reliably reproduced on the P0 (cf. Figure \ref{fig:plot11}). 

\subsection{Summary}
Following points can be taken from our work:
\begin{itemize}
\item If no explicit care is taken, networking has a huge impact on the real-timeness of embedded systems, rendering real-time guarantees invalid.
\item The empirically observed problems are largely introduced by IoT programming frameworks rather than the general architecture.
\item Transmitting command and control signals over an IP network is not feasible for tasks that need hard real-time guarantees.
\item The inflexibility of current network stack implementations and the high networking overhead suggest the reservation of a separate core for networking software. This might not be worth the cost since the previous point holds true for dual-core modules.
\item Limited access to network related tasks prevent a holistic real-time system design and cooperative scheduling between them and programmed tasks.
\end{itemize}


\section{Related Work}
\label{sec:related}

Addressing the unwanted impact of interrupts is being researched since decades.
So, to understand today's common interrupt handling mechanisms, it is helpful to understand the evolution of those mechanisms.
The first part of this section will hence briefly outline implementations that pioneered concepts which are relevant up until today.
Thereafter, we present scientific works which address the impact of \glspl{irq} on embedded real-time systems and applications.

\subsection{Basic Interrupt Handling Mechanisms}

The \textit{LynxOS} 
 \gls{rtos} has introduced the concept of running re-entrant network protocol stacks in the priority spectrum of a receiver's process.
By de-multiplexing network traffic at the \gls{nic} level, low latency for processing real-time network traffic can be guaranteed. 
LynxOS claims to be the first system built on an architecture that separates \glspl{isr} and \glspl{ist} (i.e., light weight kernel tasks) \cite{bunnell95operating}.

The \textit{SUN} (now \textit{Oracle}) \textit{Solaris} \gls{os} \cite{oracle19oracle} is another prominent example for implementing interrupt handling within the priority spectrum of regular process and thread scheduling.
Solaris distinguishes real-time, system, and time-sharing scheduling classes.
With these priority spectrums, Solaris was able to effectively isolate real-time processing from networking.


\textit{Windows Embedded Compact} (\textit{Windows CE}) \cite{microsoft13history}, is an \gls{os} family developed for handheld consumer electronics (CE) devices.
Interrupt handling is divided among the kernel and a process running all device drivers.
By running network handlers with a priority lower than the one used for real-time tasks, Windows CE effectively is able to implement isolation and call admission for incoming network activities.\\

These \glspl{os} served millions of industrial applications in practice and the influences of their concepts can still be found today.
Nevertheless, applications, their surrounding environments, and their underlying hardware have changed, where even entry-level hardware has complex performance features.
All together, these developments complicate the timing predictability and existing knowledge needs to be confirmed or refined with up-to-date numbers.

\subsection{Advanced Interrupt Handling}

Several works have identified the duality in priority spaces as still being a challenge for \glspl{rtos} and applications.

A class of approaches tries to mitigate the unwanted effects with additional hardware.
This hardware usually intercepts the \glspl{irq} between their source and the CPU.
Gomes et al. propose to extend the interrupt controller with a task-aware priority controller \cite{task_aware}.
This controller compares the priority of an incoming \gls{irq} with the priority of the currently running task and avoids that lower-priority tasks preempt higher-priority tasks.
In a proposal by Leyva-del-Foyo et al. 
a custom interrupt controller \cite{leyvadelfoyo04custom} is used to realize dynamic priorities for the \gls{irq} lines.
Additionally, all \glspl{isr} become \glspl{ist}, which enables the \gls{fpga} to unify their synchronization and scheduling.
While these solutions promise to have little overhead, the requirements of additional interrupt hardware can not be met with \gls{cots} microcontrollers.

Other works try to mitigate the unwanted effects using software-only.
In \cite{leyvadelfoyo06predictable}, Leyva-del-Foyo et al. enhance the applicability of their concept presented in \cite{leyvadelfoyo04custom} with an implementation for standard PC interrupt hardware.
Ober et al. patented a solution where the decision whether to wake an \gls{ist} or not is based on a unique priority per task and a global interrupt priority value \cite{ober10interrupt}.

Prominent work regarding the unification of priority spaces is the approach implemented in the \textit{Sloth} \gls{os} \cite{sloth, sleepy_sloth}.
Instead of using a software scheduler for threads, every control flow is designed as a thread-related system call using the hardware interrupt system.
By letting the hardware manage all control flows,
\glspl{isr} and (other) threads preempt each other in accordance with their priorities.

Although sophisticated mitigation approaches are being developed, only a few works quantify the impact of \gls{irq} loads on real-time tasks.
In \cite{regnier08evaluation}, the authors explicitly specify the \gls{irq} load caused with network packets (20 Hz) while measuring \gls{isr} latencies.
Furthermore, Regehr et al. present a collection of approximate \gls{irq} frequencies \cite{regehr05preventing}.\\ 





The research described shows that there are promising solutions for an improved latency or efficiency of interrupt handling.
However, none of the examined papers focus on the impact of \glspl{irq} on real-time tasks and do not quantify the effects directly.



\balance
\section{Conclusion}
\label{sec:conclusion}
The IoT introduces network interfaces and full IP stacks to microcontrollers running real-time applications. These open up the systems to interrupts network-triggered \glspl{irq}. We therefore analyzed the impact of network packet floods to the lateness and performance of real-time tasks on two state-of-the-art IoT \glspl{mcu}. Our results show that the execution of network stack tasks on IoT devices can pose a significant threat to real-time guarantees and that the ISR executions themselves have a similar, yet less severe impact on critically prioritized tasks in comparison to the entire packet handling.

The results have shown that a more comprehensive study with a broader range of IoT devices and test scenarios is in order. Furthermore, we will investigate mitigation techniques for NIC-generated ISR delays in real-time systems.

\section*{Acknowledgments}
We thank Martin Haug and Laurenz Mädje for their support during experiment implementations.
\bibliographystyle{IEEEtran}
\bibliography{IEEEabrv,inc/bibliography}
\end{document}